\documentclass[twocolumn%
              ,superscriptaddress%
              ,aps%
              ,pra]{revtex4-2}
\usepackage[utf8]{inputenc}
\usepackage[T1]{fontenc}
\usepackage{amsmath,amssymb}
\usepackage{braket}
\usepackage{graphicx}
\usepackage{siunitx}
\usepackage{hyperref}
\usepackage{color}
\hypersetup{colorlinks,urlcolor=black,citecolor=black,linkcolor=black,filecolor=black}

\newcommand{\e}{\mathrm{e}}
\newcommand{\tr}{\mathrm{tr}}
\newcommand{\TE}{\mathrm{TE}}
\newcommand{\TM}{\mathrm{TM}}
\newcommand{\E}{\mathrm{E}}
\newcommand{\M}{\mathrm{M}}
\newcommand{\out}{\mathrm{out}}
\newcommand{\reg}{\mathrm{reg}}

\begin{document}
\title{Switching the sign of the Casimir force between two PEMC spheres}

\author{Tanja Schoger}
\author{Gert-Ludwig Ingold}
\affiliation{Institut für Physik, Universität Augsburg, 86135 Augsburg, Germany}

\begin{abstract}
For non-reciprocal objects in vacuum, the Casimir interaction can become repulsive.
Here, we present a comprehensive study for idealized non-reciprocal materials known
as perfect electromagnetic conductors (PEMC). The system consists of two spheres
made of different PEMC materials, including the plane-sphere geometry as a particular case.
The sign of the Casimir force does not only depend on the distance between the
spheres and their geometric parameters but can be controlled by adjusting the temperature.
A repulsive Casimir interaction at small distances allows for stable equilibrium
configurations of the spheres. A sum rule previously derived for the plane-plane geometry
at zero temperature is violated in general, if at least one plane is
replaced by a sphere.
\end{abstract}

\maketitle

\section{Introduction}
Quantum and thermal fluctuations of the electromagnetic field can dominate the
interaction of neutral objects at the nano-scale through the Casimir effect.
Typically, the associated force is attractive and at submicron distances becomes
strong enough to be relevant for micro- and nano-electromechanical systems.
Resulting phenomena like stiction can then affect the functionality of such devices
\cite{Serry1998}. The possibility to switch from attractive to repulsive Casimir 
interaction through an external control parameter like the temperature is
therefore of great interest. Furthermore, if the Casimir force
changes its sign from repulsive to attractive with increasing distance between
the objects, a stable equilibrium configuration exists.

Casimir repulsion was already studied and experimentally realized in various
systems.  For dielectric objects of relative permittivity $\epsilon_1$ and
$\epsilon_2$ immersed in a liquid of permittivity $\epsilon_3$, it has been
known for a long time that repulsion can occur provided $\epsilon_1(i\xi) >
\epsilon_3(i\xi) > \epsilon_2(i\xi)$ at imaginary frequencies $\xi$
\cite{Dzyaloshinskii1961}. More recently, repulsion in such a setup was
demonstrated experimentally \cite{Munday2009}.

A repulsive Casimir force can also be realized without a medium between
objects.  Already Boyer \cite{Boyer1974} pointed out that a perfect electric
conductor and a perfect magnetic conductor  repel each other.  Since then
various systems have been studied where repulsion can occur, involving
metamaterials \cite{Yannopapas2009, Zhao2009}, topological insulators
\cite{GrushinCortijo2011, Grushin2011, Rodriguez-Lopez2011, Nie2013, 
Rodriguez-Lopez2014, Fuchs2017}, Weyl
semimetals \cite{Wilson2015, Yohei2023} or magnetoelectric materials
\cite{Genov2014, Li2021} to name but a few.

In 2018, Rode \textit{et al.}\ \cite{Rode2018} generalized the system
considered by Boyer by studying the Casimir interaction between so-called
perfect electromagnetic conductor (PEMC) plates at zero temperature. PEMC
materials interpolate between a perfect electric and magnetic conductor
\cite{Sihvola1991}.  They are a special kind of non-reciprocal medium which can
be obtained as a limit from a broader class of polarization-mixing materials,
the so-called biisotropic materials \cite{Lindell1994}.

Theoretical studies of objects embedded in a medium fulfilling the condition
mentioned above revealed the existence of a stable equilibrium position which
depends on the size of the objects as well as on temperature
\cite{Rodriguez2010a, Rodriguez2010b}. In the present paper, we extend this
approach to non-reciprocal materials in vacuum, for which, except for a study
within the pairwise summation approximation \cite{Rodriguez-Lopez2011}, only
planar geometries have been examined so far. Specifically, we will analyze the
Casimir force for two dissimilar PEMC spheres for arbitrary radii and
distances. Of particular interest is the influence of thermal fluctuations,
since the temperature can serve as an external control parameter. The sphere-plane
geometry will be included in our analysis as a limiting case.

In the following, we will provide analytical and numerical results for
the Casimir interaction between two dissimilar PEMC spheres to cover
the full range of parameters. Our numerical results are based on a plane-wave
description introduced earlier for dielectric spherical objects
\cite{Spreng2020}. For the present work, we adapted this approach to include
biisotropic materials.

This paper is organized as follows. In Sec.~\ref{sec:pemc_sphere}, we
describe the scattering of electromagnetic waves at a single biisotropic
sphere including the limiting case of a sphere consisting of a PEMC material.
Section~\ref{sec:scattering_approach} continues with a brief introduction to
the scattering approach to the Casimir interaction and its application to
spherical objects.  Next, we apply the scattering formalism to calculate the
Casimir interaction for short (Sec.~\ref{sec:short_distance}) and large
separations (Sec.~\ref{sec:large_distance}) between the PEMC spheres.
Additional analytical results for the Casimir interaction are obtained for high
temperatures in Sec.~\ref{sec:high_temp}. In Sec.~\ref{sec:sum_rule}, we
examine whether a sum rule for the Casimir force derived for the plane-plane
geometry \cite{Rode2018} carries over to the sphere-sphere setup.
Section~\ref{sec:discussion} discusses the Casimir force over the whole range
of distances, temperatures, and geometrical parameters. We investigate
the conditions for which temperature can serve as a control parameter to switch
from attraction to repulsion as well as the conditions for stable equilibrium
positions. Concluding remarks are given in Sec.~\ref{sec:conclusions} and the
Appendix contains some definitions and technical details.

\section{Scattering at a PEMC sphere}\label{sec:pemc_sphere}

The reflection operator at a single sphere is the main ingredient for
the scattering approach to the Casimir interaction which we will discuss in the
next section. We can profit from previous work \cite{Bohren1974} by starting
with a biisotropic model from which PEMC can be obtained as a limiting case.

The constitutive equations for a biisotropic material in the frequency domain are given by
\begin{equation}
\begin{pmatrix}
\mathbf{D} \\ \mathbf{B}
\end{pmatrix} = 
\begin{pmatrix}
\epsilon & \alpha\\
\beta & \mu
\end{pmatrix}
\begin{pmatrix}
\mathbf{E} \\ \mathbf{H}
\end{pmatrix}
\label{eq:constitutive_eq}
\end{equation}
with the permittivity $\epsilon$ and the permeability $\mu$.
The biisotropy parameters $\alpha$ and $\beta$ account for the magneto-to-electric
and electro-to-magnetic coupling, respectively. All four parameters are scalar
functions of the frequency. If the Onsager reciprocal relation $\alpha =
-\beta^*$ is violated, the material is non-reciprocal (see e.g.
\cite{Caloz2018} for a review).  The constitutive equations above are, for example,
used to describe topological insulators (see e.g. \cite{Silva2022} for a
review). 

The reflection operator $\mathcal{R}$ of a spherical object is usually
expressed in a spherical-wave basis $|\ell, m, P, s \rangle$.  Each multipole
mode is defined by its angular momentum $\ell = 1, 2, \ldots $ with $m = -\ell,
\ldots 0,\ldots \ell$, the polarization $P$, which is either electric ($\E$) or
magnetic ($\M$) and a parameter $s$ which describes an incoming ($\reg$) or
outgoing ($\out$) wave with respect to the sphere center.  Within the
$2\times2$-dimensional polarization subspace of the spherical-wave basis, the
matrix elements of the reflection operator for a biisotropic sphere can be
expressed as 
\begin{equation}
\langle P | \mathcal{R} | P' \rangle
=-i^{P'-P}(\mathbf{R})_{P, P'} \,,
\label{eq:R_multipole}
\end{equation}
where we associate $P=1$ $(P=2)$ to E (M) polarized modes and the matrix
$\mathbf{R}$ contains the Mie coefficients which for biisotropic materials
were derived in \cite{Bohren1974}. As the constitutive equations used there
differ from \eqref{eq:constitutive_eq} and we consider imaginary frequencies
instead of real frequencies, we provide explicit expressions
for the Mie coefficients in Appendix \ref{sec:mie_coef}.

By choosing the material parameters in the constitutive equation
\eqref{eq:constitutive_eq} as
\begin{equation}
\alpha = \beta = q, \quad
\epsilon = q\cot(\theta), \quad
\mu = q \tan(\theta)\,,
\label{eq:def_param_pemc}
\end{equation}
one obtains PEMC by taking the limit $q \rightarrow \infty$. This class of
materials is parametrized by the angle $\theta$ taking values between 0 and
$\pi/2$. The two boundary cases correspond to the perfect electric (PEC) and
perfect magnetic (PMC) conductor, respectively \cite{Lindell1994}. According to
the choice of $\alpha$ and $\beta$ in \eqref{eq:def_param_pemc}, PEMC are
non-reciprocal.

It can be shown that the reflection matrix in \eqref{eq:R_multipole} for a PEMC
sphere can be expressed as
\begin{equation}
\mathbf{R}_\mathrm{PEMC} = \mathbf{D}\mathbf{R}_\mathrm{PEC}\mathbf{D}^{-1}\,,
\label{eq:R_PEMC}
\end{equation}
where $\mathbf{R}_\mathrm{PEC}$ is the reflection matrix for a perfect electric
conductor while the duality transformation matrix
\begin{equation}
\mathbf{D} = \begin{pmatrix}
\cos(\theta) & \sin(\theta) \\
-\sin(\theta) & \cos(\theta)
\end{pmatrix}
\label{eq:def_D}
\end{equation}
depends on the material parameter $\theta$ \cite{Buhmann2009}.
Explicit expressions are given in \eqref{eq:al_pemc}--\eqref{eq:dl_pemc}.

Clearly, PEMC constitute an idealization of real non-reciprocal materials. For
a theoretical study, however, they allow to characterize the material
properties of the system by the single parameter $\theta$. We can thus analyze
the interplay between geometry, temperature and material properties of the system.
Furthermore, they
may provide a guiding line for future experiments aiming at realizing Casimir
repulsion by making use of non-reciprocal materials. Moreover, the PEMC model
allows us to derive analytical results for the Casimir interaction, as can be seen
in the following sections. 

\section{Scattering approach to the Casimir interaction}
\label{sec:scattering_approach}

The scattering approach to the Casimir effect was originally developed for
objects consisting of reciprocal materials \cite{Lambrecht2006, Emig2007}.
A recent general study confirmed that in thermal equilibrium, the scattering approach
holds even for non-reciprocal objects \cite{Gelbwaser2022}. For two non-reciprocal
plates, this was already shown in \cite{Fuchs2017}.  

\begin{figure}
 \includegraphics[width=0.35\textwidth]{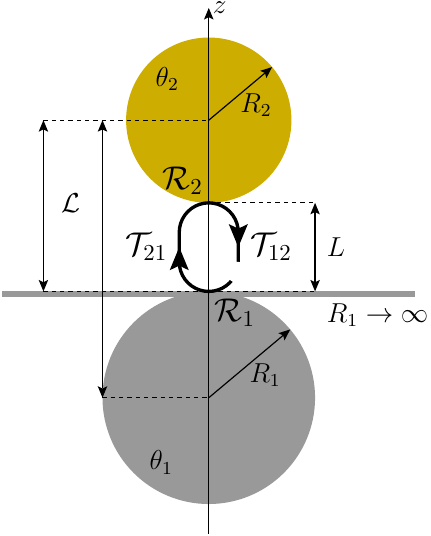}
 \caption{Scattering geometry consisting of two spheres or, in the limit
          $R_1\to\infty$, a sphere in front of a plane placed along the $z$-axis
          with a surface-to-surface distance $L$. In the sphere-sphere geometry, 
          we also define the distance $\mathcal{L} = L + R_1 + R_2$ and for
          the sphere-plane setup we will use $\mathcal{L} = L + R_2$. 
          The objects are made of PEMC with
          parameters $\theta_1$ and $\theta_2$. The round-trip operator
          \eqref{eq:roundtripoperator} is indicated by the black loop in
          the gap.}
\label{fig:geometry}
\end{figure}

In our calculations, we will evaluate as the primary quantity the Casimir
free energy $\mathcal{F}$ from which the Casimir force $F$ can be obtained
by taking the derivative with respect to the surface-to-surface
distance $L$ between the objects according to
\begin{equation}
  F = - \frac{\partial\mathcal{F}}{\partial L}\,.
  \label{eq:force}
\end{equation}
At finite temperatures, the Casimir free energy can be expressed as a
sum over Matsubara frequencies $\xi_n = 2\pi k_\mathrm{B}T n/\hbar$ 
\begin{equation}
\mathcal{F} = \frac{k_\mathrm{B} T}{2}\sum_{n=-\infty}^\infty f_{|n|}
\label{eq:full_freeenergy}
\end{equation}
with
\begin{equation}
f_n = \log\det(1-\mathcal{M}(i\xi_n))
= - \sum_{r=1}^\infty \frac{\mathrm{tr}\mathcal{M}^r(i\xi_n)}{r}\,.
\label{eq:def_Fn}
\end{equation}
The round-trip operator
\begin{equation}
\mathcal{M} = \mathcal{T}_{12}\mathcal{R}_{2}\mathcal{T}_{21}\mathcal{R}_1
\label{eq:roundtripoperator}
\end{equation}
describes the scattering process of the electromagnetic field between the two
objects. The operator $\mathcal{R}$ accounts for the reflection at an object
and the operator $\mathcal{T}$ translates the electromagnetic field from the
reference frame of one object to the other (see Fig.~\ref{fig:geometry}).

In the second equality in \eqref{eq:def_Fn} we made use of Jacobi's formula and
the Mercator series. The summation index $r$ accounts for the number of
round-trips of the electromagnetic field between the two objects. The
single-round-trip contribution is thus obtained by retaining only the term with
$r=1$. While the round-trip expansion is particularly useful to derive
analytical results, numerical approaches are usually based on evaluating
the determinant in \eqref{eq:def_Fn}.

At zero temperature, the Matsubara sum turns into an integral over 
imaginary frequencies $\omega = i \xi$. In the opposite limit of
high temperatures, the free energy
\begin{equation}
\mathcal{F}_T = \frac{k_\mathrm{B}T}{2}f_0
\label{eq:FT}
\end{equation}
depends exclusively on the contribution of the Matsubara frequency $\xi_0$.
This expression is proportional to the temperature and of purely
entropic origin.

In order to calculate the Casimir free energy, we evaluate the trace in
\eqref{eq:def_Fn} in an appropriate basis set, depending on the geometry of the
scattering problem. We will specifically consider the setup depicted in
Fig.~\ref{fig:geometry} consisting of two spherical particles with radii $R_1$
and $R_2$ at a surface-to-surface distance $L$ and made of PEMC with parameters
$\theta_1$ and $\theta_2$. The sphere centers are located on the $z$-axis and
serve as origins of the reference frames in which the reflection of
electromagnetic waves is described. Their distance is given by $\mathcal{L} = L
+ R_1 + R_2$. We will later make use of the aspect ratio
\begin{equation}
  x = \frac{L}{R_\text{eff}}
  \label{eq:def_x}
\end{equation}
with the effective radius
\begin{equation}
  R_\text{eff} = \frac{R_1R_2}{R_1+R_2}\,.
  \label{eq:effective_radius}
\end{equation}
In the limit $R_1\to\infty$, the sphere-sphere geometry turns into the
sphere-plane geometry. The transition from equally sized spheres to a sphere
in front of a plane can be described by the dimensionless parameter
\begin{equation}
  u = \frac{R_1R_2}{(R_1+R_2)^2}
  \label{eq:def_u}
\end{equation}
which takes values between $0$ in the case of the sphere-plane geometry and
$1/4$ for two spheres of equal radii.

While the reflection operator matrix elements are conveniently expressed in
the spherical-wave basis introduced in Sec.~\ref{sec:pemc_sphere}, a plane-wave
description is better suited for the translation operator. For the plane-wave
basis, we employ the angular spectral representation \cite{Nieto-Vesperinas2006}.
Omitting the imaginary frequency $\xi$, which is constant during the whole
round-trip, we denote the basis by $|\mathbf{k}, p, \pm \rangle$, where 
$\mathbf{k} = k(\cos(\varphi), \sin(\varphi))$ is the transversal part 
of the wave vector $\mathbf{K}$. The
dispersion relation can then be expressed as $\xi^2=c^2(\mathbf{k}^2-\kappa^2)$,
where $\kappa$ is the imaginary wave vector component in $z$-direction. The
polarization $p$ of the plane wave is defined with respect to a plane perpendicular
to the $z$-axis and can be transverse magnetic ($\TM$) or transverse
electric ($\TE$).  The signs $\pm$ refer to the direction of propagation along
the $z$-axis.  Within the plane-wave basis, the translation operator is
diagonal with the matrix elements
\begin{equation}
\langle \mathbf{k}, p, \pm| \mathcal{T}| \mathbf{k}', p', \pm \rangle = 
e^{-\kappa \mathcal{L}} \delta(\mathbf{k}-\mathbf{k}')\delta_{p, p'}.
\label{eq:translation}
\end{equation}
The trace over the $r$-fold round-trip operator \eqref{eq:def_Fn} in the plane-wave basis
is thus given by \cite{Spreng2018}
\begin{equation}
\begin{aligned}
\tr\mathcal{M}^r &= \sum_{p_1,\ldots, p_{2r}} 
\int\frac{\mathrm{d}\mathbf{k}_1\ldots 
\mathrm{d}\mathbf{k}_{2r}}{(2\pi)^{4r}} 
\prod_{j=1}^{r} \e^{-\kappa_{2j}\mathcal{L}} \e^{-\kappa_{2j-1}\mathcal{L}}\\
&\hspace{4em}
\times \langle \mathbf{k}_{2j+1}, p_{2j+1}, - \vert
\mathcal{R}_2 \vert \mathbf{k}_{2j}, p_{2j}, + \rangle \\
&\hspace{4em}
\times\langle \mathbf{k}_{2j}, p_{2j}, + \vert
\mathcal{R}_1 \vert \mathbf{k}_{2j-1}, p_{2j-1}, - \rangle\,,
\end{aligned}
\label{eq:def_trMr}
\end{equation}
where cyclic indices $2r+1 \equiv 1$ are used to account for the trace. 

Within the plane-wave basis, the reflection operator can be determined by
means of a basis change
\begin{equation}
\langle p | \mathcal{R} | p' \rangle = \sum_{P, P' = \E, \M}
\langle p | P \rangle \langle P | \mathcal{R} | P' \rangle
\langle P' | p' \rangle 
\label{eq:basis_change}
\end{equation}
and making use of the reflection matrix elements in the spherical-wave
basis discussed in Sec.~\ref{sec:pemc_sphere} and Appendix~\ref{sec:mie_coef}.
For simplicity, we omitted in \eqref{eq:basis_change} all indices of the bases,
except for the polarization. The basis transformation coefficients
$\langle p | P \rangle$ and $\langle P | p \rangle$ can be found in 
\cite{Messina2015}. Together with \eqref{eq:R_multipole} we thus obtain for a
biisotropic sphere \cite{Ingold2022}
\begin{equation}
\begin{aligned}
\langle\mathbf{k}, p\vert \mathcal{R}\vert \mathbf{k}', p'\rangle =
\frac{2\pi c}{\xi \kappa}&\bigg[A S_{p, p'}
	+(-1)^{p+p'} B S_{\bar{p},\bar{p}'}\\
& - (-1)^{p} C S_{\bar{p}, p'}
	+ (-1)^{p'} D S_{p, \bar{p}'}\bigg]\,.
\end{aligned}
\label{eq:refl_coef_planewave}
\end{equation}
In the notation above, we associate $\bar{p}$ with the opposite polarization to $p$ and we set 
$p=1$ ($p=2$) for $\TM$ ($\TE$) polarized waves. 
The coefficients $A,\, B,\, C$ and $D$ account for the polarization transformations 
and are given in \cite{Spreng2018}. They depend on the incoming and outgoing wave vectors 
$A = A(\mathbf{K}, \mathbf{K'})$. 
$S_{p, p'}$ are the elements of the amplitude scattering matrix \cite{Bohren1974}
and are reproduced in \eqref{eq:scattering_ampl} for imaginary frequencies and our
definitions of the reflections matrix elements.  

As indicated in \eqref{eq:FT}, the high-temperature limit of
the Casimir free energy is determined by the zero-frequency contribution, for
which the amplitude scattering matrix simplifies. The polarization changing
coefficients $B, C$ and $D$ in \eqref{eq:refl_coef_planewave} vanish in this
limit, while $A$ tends to one \cite{Spreng2018}. Furthermore, following the
argument given in Appendix~\ref{sec:mie_coef}, even in the plane-wave basis
the leading order of the reflection coefficients for a PEMC sphere in the
low-frequency limit are obtained by a duality transformation
\begin{equation}
\langle p| \mathcal{R}| p' \rangle = \frac{2\pi c}{\xi k}
\left(\mathbf{D}\mathbf{S}_\text{PEC}\mathbf{D}^{-1}\right)_{p, p'}\,.
\label{eq:scat_zerofreq}
\end{equation}
Here, $\mathbf{S}_\text{PEC}$ defines the amplitude scattering matrix for a perfect
reflector in the low-frequency limit \eqref{eq:scat_ampl_zerofreq}. 
This simplification at very low frequencies will allow for analytical
calculations of the Casimir interaction in the high-temperature regime.  

\section{Short-distance asymptotics}\label{sec:short_distance}

If the sphere radii are large compared to the surface-to-surface distance,
$R_1, R_2 \gg L$, the Casimir force is commonly treated within the
proximity-force approximation (PFA) \cite{Derjaguin1934}.  This approximation
treats the Casimir interaction between the spheres in close vicinity as the
interaction between parallel plate segments. As demonstrated in
\cite{Ingold2022}, the PFA result for two PEMC spheres can be obtained from an
asymptotic expansion of the reflection coefficients for large radii. The
round-trip decomposition of the Matsubara frequency contributions
(\ref{eq:def_Fn}) was found as
\begin{equation}
f_{n, \text{PFA}} = - R_\text{eff} \int \frac{\mathrm{d}\mathbf{k}}{2\pi \kappa} 
\mathrm{Re}\left[\mathrm{Li}_2(e^{2i\delta -2 \kappa L})\right]\,,
\end{equation}
where $\mathrm{Li}_{n}(z)$ is the polylogarithm of order $n$.

The integral over $\mathbf{k}$ can be calculated explicitly and yields
\begin{equation}
f_{n, \text{PFA}} = - \frac{1}{2x}
\mathrm{Re}\left[\mathrm{Li}_3( e^{2i\delta -2 n\tau})\right]\,,
\label{eq:Fn_PFA}
\end{equation}
where $x$ is the aspect ratio defined in (\ref{eq:def_x}) and
$\tau=2\pi L/\lambda_T$ is a dimensionless temperature with the thermal 
wavelength $\lambda_T = \hbar c/k_\mathrm{B} T$. The coefficients (\ref{eq:Fn_PFA})
and, as a consequence, the Casimir energy depend 
only on the difference of the PEMC parameters $\theta_{1,2}$ of the two
spheres
\begin{equation}
\delta = |\theta_2 - \theta_1|\,.
\end{equation}
The limiting values $\delta=0$ and $\pi/2$ correspond to two identical PEMC
spheres and to a PEC and a PMC sphere, respectively. The nature of the Casimir
force thus changes from attractive for two symmetric particles
\cite{Kenneth2006} to repulsive with increasing $\delta$.  

In the zero-temperature limit, we already confirmed in \cite{Ingold2022} that our PFA result 
agrees with the one obtained by Rode et al. \cite{Rode2018} 
and earlier ones \cite{Asorey2013, Feng2014} obtained for a scalar field with 
pseudo-periodic boundary conditions
\begin{equation}
\mathcal{E}_\mathrm{PFA} = - \frac{\hbar c}{720\pi L}\frac{1}{x}
	\left[\pi^4-30\delta^2(\pi-\delta)^2\right]\,.
\label{eq:E_PFA}
\end{equation}
This expression for the Casimir energy changes sign between $\delta=0$
and $\pi/2$ and according to \eqref{eq:force} the Casimir force
changes sign as well. The critical value of $\delta$ at which
the force vanishes is given by
\begin{equation}
  \delta_\mathrm{crit}^{T=0} = \left(1-\sqrt{1-2\sqrt{\frac{2}{15}}}\right)\frac{\pi}{2}
  = 0.961\dots\frac{\pi}{4}\,.
\label{eq:dc_PFA_T0}
\end{equation}
Here and in the following, it is convenient to express critical angles
in units of the central angle $\pi/4$ corresponding to a situation half-way
between equal materials and a PEC-PMC setup.

The force thus changes its sign at $\delta_\mathrm{crit}$ and is attractive
for systems with $\delta < \delta_\mathrm{crit}$ while being repulsive for
$\delta > \delta_\mathrm{crit}$.  In the following, we are going to determine
the temperature and distance corrections to the PFA result \eqref{eq:E_PFA} and
analyze how $\delta_\mathrm{crit}$ changes when we increase the temperature or
the separation between the spheres. 

\subsection{Temperature corrections to PFA}

We will now examine the temperature corrections to the critical angle
introduced in \eqref{eq:dc_PFA_T0}.  We thus consider the case 
$L \ll \lambda_T \ll R_\mathrm{eff}$ for $\delta > 0$,
where the temperature corrections for $\tau \ll 1$
can be calculated by applying the Mellin transform of the exponential in \eqref{eq:Fn_PFA}. After
performing the sum over the Matsubara frequencies \eqref{eq:full_freeenergy}, we obtain 
\begin{widetext}
\begin{equation}
\mathcal{F}_\text{PFA} = - \frac{k_\mathrm{B} T}{4x}\left[
\mathrm{Re}\left[\mathrm{Li}_3(e^{2i \delta})\right]
 + 2\int_{c-i \infty}^{c+i\infty}  \frac{\mathrm{d}s}{2\pi i}
\Gamma(s) \zeta(s)\mathrm{Re}\left[ \mathrm{Li}_{s+3}(e^{2i\delta})\right]
(2 \tau)^{-s}
\right]\,,
 \label{eq:PFA_free_energy}
\end{equation}
where $\Gamma(s)$ and $\zeta(s)$ refer to the gamma function and the Riemann
zeta function, respectively. The integration contour in the complex plane
is chosen such that $c > 1$. 
\end{widetext}
For $\delta > 0$, the integrand has single poles at $s = 1, 0$ and $-2n-1$ with $n = 0, 1, 2, \ldots$
The pole at $s=1$ accounts for the zero-temperature result \eqref{eq:E_PFA}. 
The leading temperature corrections then arise from the poles at $s=0, -1$ and $-3$. 
Applying Cauchy's theorem and evaluating the residues, we find for the low-temperature expansion
\begin{equation}
\begin{aligned}
\Delta\mathcal{F}_\text{PFA} 
&= \mathcal{F}_\mathrm{PFA}- \mathcal{E}_\mathrm{PFA} \\
&= -\frac{\hbar c }{720\pi L x}\left[
5\left(\pi^2 - 6\delta (\pi - \delta)\right) \tau^2
 + \tau^4 + \mathcal{O}(\tau^6)
\right]\,, 
\end{aligned}
\label{eq:PFA_temp_corr}
\end{equation}
where we used the representation of the polylogarithm in terms of the Bernoulli polynomials \cite[24.8.3]{DLMF}
\begin{equation}
  \mathrm{Li}_n(e^{2\pi i z})+ (-1)^n \mathrm{Li}_n(e^{-2\pi i z}) = - \frac{(2\pi i)^n}{n!} B_n(z)\,. 
\end{equation}
In the limit $\delta = \pi/2$, our result agrees with the one in \cite{Teo2012}.

\begin{figure}
\includegraphics[width=0.4\textwidth]{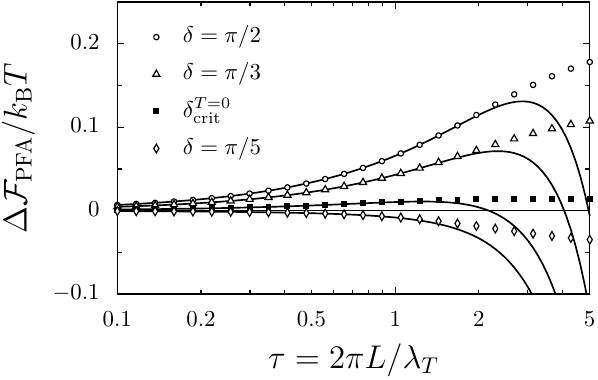}
\caption{Temperature corrections to the PFA result \eqref{eq:E_PFA} for the
         Casimir energy in units of $k_\mathrm{B}T$ for various values of $\delta$. 
         The symbols depict the numerically exact values and the solid lines
         represent the low-temperature expansion given in \eqref{eq:PFA_temp_corr}.}
\label{fig:temp_corr_pfa}
\end{figure}

The range of validity of our result can be tested by comparing the
low-temperature expansion with the numerically evaluated Matsubara sum. Fig.~\ref{fig:temp_corr_pfa}
shows the numerically calculated temperature corrections represented by dots
and the asymptotic expansion \eqref{eq:PFA_temp_corr} represented by lines
as function of the effective temperature $\tau = 2\pi L/\lambda_T$ and for different
values of $\delta$.  With decreasing temperature, the asymptotic expansion
converges towards the numerically exact results as expected.

Already the results displayed in Fig.~\ref{fig:temp_corr_pfa} indicate that
$\delta_\text{crit}$ depends on temperature. In fact, the low-temperature corrections
for $\delta=\delta_\text{crit}^{T=0}$ (filled squares) become positive for
non-vanishing temperatures, suggesting that the value of the critical angle
decreases with increasing temperature. This expectation is further corroborated
by the high-temperature limit of the free energy in the PFA limit given by the
first term in \eqref{eq:PFA_free_energy} as
\begin{equation}
\mathcal{F}_{T, \text{PFA}} = -\frac{k_\mathrm{B} T}{4x}
\mathrm{Re}\left[\mathrm{Li}_3(e^{2i \delta})\right]
\label{eq:freeenergy_hightemp}
\end{equation}
which is in agreement with results found in the literature for
$\delta=0$ and $\delta = \pi/2$ \cite{daSilva2001, Lim2009}. 
The high-temperature expression implies
that the Casimir force vanishes at a critical value 
\begin{equation}
\delta_\mathrm{crit}^{T\rightarrow \infty} = 0.923 \ldots \frac{\pi}{4}\,,
\label{eq:dc_PFA_HT}
\end{equation}
i.e., at a value below the zero-temperature critical value \eqref{eq:dc_PFA_T0}.

\begin{figure}
\centering
\includegraphics[width=0.4\textwidth]{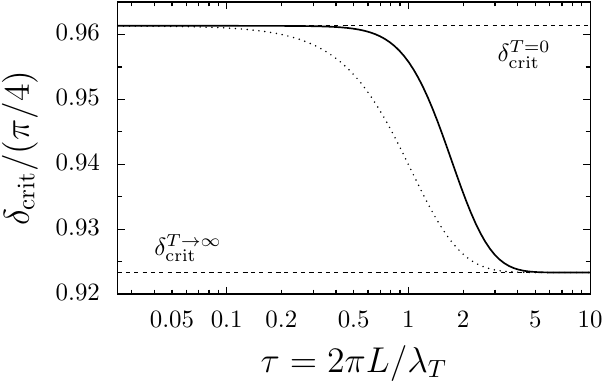}
\caption{The temperature dependence of $\delta_\text{crit}$ in the PFA regime
         is shown by the solid line. For comparison, the value of $\delta$
         for which the free energy vanishes, is depicted by a dotted line.
         The dashed lines indicate the low-temperature value \eqref{eq:dc_PFA_T0}
         and the high-temperature value \eqref{eq:dc_PFA_HT} of the
         critical angle.}
\label{fig:pfa_temperature_corrections}
\end{figure}

A numerical evaluation of the temperature dependence of $\delta_\text{crit}$
yields the solid line in Fig.~\ref{fig:pfa_temperature_corrections}
showing that $\delta_\text{crit}$ decreases monotonically with increasing
temperature from the zero-temperature value \eqref{eq:dc_PFA_T0} to the
high-temperature value \eqref{eq:dc_PFA_HT}. The dotted line indicates
the value of $\delta$ where Casimir free energy vanishes. As the expressions
\eqref{eq:E_PFA} and \eqref{eq:freeenergy_hightemp} for the 
Casimir free energy at zero temperature and high temperatures, respectively,
factorize into contributions depending on $L$ and $\delta$ separately,
the Casimir free energy and the Casimir force vanish at the same value
of $\delta$ in these two limits. This no longer holds at intermediate temperature,
as can be seen from the low-temperature expansion \eqref{eq:PFA_temp_corr}
where the dimensionless temperature $\tau$ introduces an additional
dependence on $L$. 

The results for the temperature dependence of $\delta_\text{crit}$ displayed in
Fig.~\ref{fig:pfa_temperature_corrections} imply that if we choose a system
which can be described by a parameter in between $\delta_\mathrm{crit}^{T=0}$
and $\delta_\mathrm{crit}^{T\to\infty}$, the Casimir force in
the PFA regime will change from attractive to repulsive with increasing
temperature. The transition occurs at temperatures around $k_\mathrm{B}T
\approx 0.2 \hbar c/L$, which at room temperature corresponds to a distance of
about $L=\qty{1.5}{\micro\metre}$. The enhancement of Casimir repulsion
due to thermal fluctuations was also predicted for systems with
metallic-based metamaterials \cite{Rosa2008} or in magnetodielectric
systems \cite{Shelden2023}.  In these cases, the repulsion
originates from the contribution of the zero-frequency TE modes. 

\subsection{Geometrical corrections to PFA} \label{sec:pfa_corrections}
Going beyond PFA at zero temperature involves taking the leading corrections of an asymptotic expansion of 
\eqref{eq:def_trMr} for $x= L/R_\mathrm{eff} \ll 1$ into account. A detailed analysis can be found in \cite{Ingold2022}
where the leading geometrical corrections $\Delta \mathcal{E}_\mathrm{PFA} = \mathcal{E}- \mathcal{E}_\mathrm{PFA}$
are shown to read
\begin{equation}
\begin{aligned}
\Delta \mathcal{E}_\text{PFA} \approx \frac{\hbar c}{720\pi L}&
\left[20\left(\pi^2-6\delta(\pi-\delta)\right)\right.
\\
& \left.-\frac{1-3u}{3}
\left(\pi^4-30\delta^2(\pi-\delta)^2\right)
\right]\,.
\end{aligned}
\label{eq:beyond_pfa}
\end{equation}
The geometrical corrections depend on the parameter $u$ introduced in
\eqref{eq:def_u} which takes values between 0 and $1/4$ corresponding to the
sphere-plane geometry and a setup of two equally-sized spheres, respectively.
The $u$-dependence indicates that with increasing separation, curvature
effects become more important. 

\begin{figure}
\includegraphics[width=0.4\textwidth]{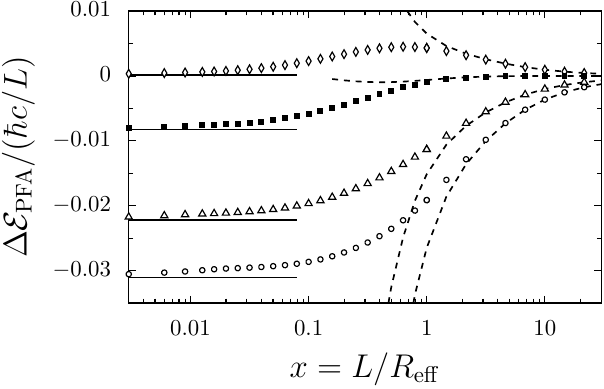}
\caption{Geometrical correction to the zero-temperature PFA result \eqref{eq:E_PFA}
         for a sphere-plane setup, \textit{i.e.}, $u=0$, as a function of the
         aspect ratio \eqref{eq:def_x} for the values of $\delta$
         used in Fig.~\ref{fig:temp_corr_pfa} and indicated by the same
         symbols. The solid lines refer to the values obtained from 
         \eqref{eq:beyond_pfa} while the dashed lines represent
         the dipole-plane approximation \eqref{eq:F_dip_plane}
         valid for large distances.}
\label{fig:beyond_pfa}
\end{figure}

In order to verify that the leading correction to the PFA result is correct, we
compare it with exact numerical results, which was not done in
Ref.~\onlinecite{Ingold2022}. Fig.~\ref{fig:beyond_pfa} depicts the numerically
calculated geometrical corrections over the whole range of aspect ratios $x$
for the sphere-plane geometry, \textit{i.e.}, $u=0$. The four data sets
correspond to the values of $\delta$ used in Fig.~\ref{fig:temp_corr_pfa}
ranging from $\pi/5$ to $\pi/2$ and indicated here by the same symbols. As
expected, for small aspect ratios, the results converge towards the values
given by \eqref{eq:beyond_pfa} and depicted by solid lines. In contrast, for
large aspect ratios, the results approach the dashed lines representing the
energy for a dipole-plane setup as given below by \eqref{eq:F_dip_plane} for
$\tilde{\tau}\to0$.

The geometrical corrections presented in Fig.~\ref{fig:beyond_pfa} let us expect that the
critical angle $\delta_\text{crit}$ where the Casimir force changes its sign
not only depends on the temperature but also on the aspect ratio. This
dependence will be discussed in more detail in Sec.~\ref{sec:discussion}
by making use of the data presented in Fig.~\ref{fig:xvsdelta}.

\section{Long-range asymptotics}
\label{sec:large_distance}

If the two objects are sufficiently far apart from each other, the exponential
decay of the translation matrix element \eqref{eq:translation} implies that the
main contribution to the free energy is given by the single-round-trip
expression obtained from \eqref{eq:def_Fn} by retaining only the term with
$r=1$. For $L\gg R_\text{eff}$, \textit{i.e.}, for distances much larger than
the radius of the smaller sphere, it is even sufficient to restrict the
single-round-trip result to the dipole contribution $\ell=1$. Then, the
trace over the round-trip operator in the multipole basis is given by
\begin{equation}
\tr\mathcal{M} \approx   \sum_{P = \E, \M } \sum_{m= -1}^1 
\langle 1, m, P | \mathcal{M} | 1, m, P \rangle\,.
\label{eq:F_dipole}
\end{equation}

The evaluation of the coefficients depends on the geometry under examination.
Both, the sphere-sphere and sphere-plane geometry have in common, that one
object is given by a sphere. Separating the reflection operator of the sphere
from the rest of the round-trip operator thus leads to
\begin{equation}
\langle   m, P | \mathcal{M}| m, P \rangle = \sum_{P' = \E, \M}  U_{P, P'}^{(m)} \langle   m, P'| \mathcal{R}_1 | m , P\rangle
\end{equation}
where, for simplicity, we drop the multipole moment from the states as in 
this section it always takes the value $\ell=1$. Furthermore, we made use
of the fact that the reflection matrix in the multipole basis is diagonal
with respect to $\ell$ and $m$.

The matrix elements of the reflection operator for a PEMC sphere are
given by \eqref{eq:R_multipole}, where the reflection matrix \eqref{eq:R_PEMC} in the dipole limit yields
\begin{equation}
\mathbf{R}_\mathrm{PEMC} \approx \frac{1}{6}\left(\frac{\xi R}{c}\right)^3
\begin{pmatrix}
1 + 3\cos(2\theta_1) & 3\sin(2\theta_1) \\
3\sin(2\theta_1) & 1 - 3\cos(2\theta_1)
\end{pmatrix}\,.
\label{eq:R_dipole}
\end{equation}
The matrix elements 
\begin{equation}
U_{P, P'}^{(m)} = \langle m, P| \mathcal{T}_{12}\mathcal{R}_2\mathcal{T}_{21}|m, P'\rangle
\label{eq:U_PP}
\end{equation}
account for the translation operators and the remaining reflection operator in
the multipole basis. In order to specify the matrix elements \eqref{eq:U_PP},
we need to distinguish between the sphere-sphere and sphere-plane setup.

\subsection{Dipole-dipole limit}

For two spheres with radii $R_1$, $R_2$ much smaller than the
surface-to-surface distance $L$, we can employ the dipole approximation also
for the larger sphere. The matrix elements \eqref{eq:U_PP} can thus be written as
\begin{equation}
\begin{aligned}
U_{P, P'}^{(m)} = \sum_{P'', P'''}
&\langle m, P|\mathcal{T}_{12}|m, P''\rangle \langle m, P''|\mathcal{R}_{2}|m, P'''\rangle\\
                & \times\langle m, P'''|\mathcal{T}_{21}|m, P' \rangle\,.
\end{aligned}
\end{equation}
The eigenvalue $m$ of the $z$-component of the angular momentum is conserved
because the translations take place along the symmetry axis of the setup.
The matrix elements for the reflection operator are defined in \eqref{eq:R_dipole} and 
the matrix elements of the translation operator in the spherical-wave basis 
can be found in Eqs.~(38)--(40) of Ref.~\onlinecite{IngoldUmrath2015}. 

After performing the sum over the Matsubara frequencies in \eqref{eq:F_dipole}, we obtain
\begin{widetext}
\begin{equation}
\mathcal{F}_\text{dip-dip} = - \frac{\hbar c}{2\pi \mathcal{L}}
\left(\frac{R_1 R_2}{\mathcal{L}^2}\right)^3 \left[
\cos^2(\delta)\left( f_{P, P}(\tilde\tau) + f_{P, \bar{P}}(\tilde\tau)\right)
- \sin^2(\delta)\left( \frac{4}{5} f_{P, P}(\tilde\tau) + \frac{5}{4} f_{P, \bar{P}}(\tilde\tau)\right)
\right]
\label{eq:F_dip_dip}
\end{equation}
\end{widetext}
where $f_{P, P} (\tilde\tau)$ and $f_{P,\bar{P}}(\tilde\tau)$ are analytic functions of the
effective temperature $\tilde\tau = 2\pi\mathcal{L}/\lambda_T$ which account for
channels conserving or changing polarization upon translation, respectively.
Note that the dimensionless temperatures $\tau$ and $\tilde\tau$ use different distances
$L$ and $\mathcal{L}$, respectively.
These functions can be found in Ref.~\onlinecite{IngoldUmrath2015} and are reproduced in
\eqref{eq:dipole_dipole} for convenience.
In the zero-temperature limit $\tilde\tau \rightarrow 0$, the Casimir energy is given by
\begin{equation}
\mathcal{E}_\text{dip-dip} = - \frac{\hbar c}{16\pi \mathcal{L}} 
\left(\frac{R_1 R_2}{\mathcal{L}^2}\right)^3
[8+135\cos(2\delta)]\,.
\label{eq:E_dipole-dipole}
\end{equation}
For $\delta = 0$ and $\pi/2$ the expressions agree with the results obtained in
Refs.~\onlinecite{Rodriguez-Lopez2011} and \onlinecite{Boyer1974}, respectively. 

In the previous section, it was already suspected that $\delta_\mathrm{crit}$
changes with the distance between the spheres. The upper bound for the critical
angle in the zero-temperature limit can now be determined 
from \eqref{eq:E_dipole-dipole} and yields
\begin{equation}
\delta_\text{crit}^{T=0} 
= \frac{1}{2}\arccos\left(-\frac{8}{135}\right)
=  1.037\ldots\frac{\pi}{4}\,.
\label{eq:dc_T0_dipole}
\end{equation}
The critical angle, where the force vanishes, thus increases with distance 
as can be seen by comparing \eqref{eq:dc_T0_dipole} with \eqref{eq:dc_PFA_T0} for short separations.
A more complete picture will be given in Fig.~\ref{fig:xvsdelta} and
discussed in Sec.~\ref{sec:discussion}.

In the high-temperature limit $\tilde\tau \rightarrow \infty$, the contribution of the
polarization mixing channels $f_{P, \bar{P}}$ vanishes and the free energy yields 
\begin{equation}
\mathcal{F}_{T, \text{dip-dip}} = - \frac{3k_\mathrm{B}T}{8} 
\left(\frac{R_1 R_2}{\mathcal{L}^2}\right)^3 
[1 + 9 \cos(2\delta)]
\end{equation}
which also agrees with a result in the literature for $\delta =0$ \cite{Rodriguez-Lopez2011}. 
The magnetoelectric effect, responsible for the repulsion, reduces due to the vanishing polarization mixing channels. 
The critical angle is thus shifted towards a larger value compared to the zero-temperature case
\begin{equation}
\delta_\text{crit}^{T\rightarrow\infty}
= \frac{1}{2}\arccos\left(-\frac{1}{9}\right)
 = 1.070\ldots\frac{\pi}{4}\,.
\label{eq:dc_ht_dipole}
\end{equation}
Compared with the high-temperature result for short distances \eqref{eq:dc_PFA_HT}, one finds
that the critical angle increases as the separation between the spheres
grows. The distance dependence is thus similar to the one found in the
low-temperature regime.

\begin{figure}
\centering
\includegraphics[width=0.4\textwidth]{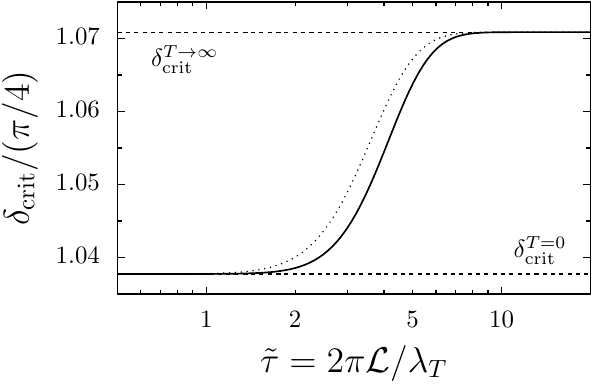}
\caption{The critical angle $\delta_\text{crit}$ as function of the temperature is shown
         as solid line in the limit of large distance between the spheres.
         The dotted line refers to the values of $\delta$ where the free energy
         changes its sign and the dashed lines indicate the low-temperature
         value \eqref{eq:dc_T0_dipole} and high-temperature value
         \eqref{eq:dc_ht_dipole} of the critical angle.}
\label{fig:dipole-dipole}
\end{figure}

As can be seen from the solid line in Fig.~\ref{fig:dipole-dipole}, the
critical angle $\delta_\text{crit}$ increases monotonically with increasing
temperature, which is consistent with the growth of $\delta_\mathrm{crit}$ with
larger distances between the spheres.  For systems with $\delta$ in between the
two limiting cases \eqref{eq:dc_T0_dipole} and \eqref{eq:dc_ht_dipole}, the
force changes from repulsion to attraction upon increasing the temperature. The
transition occurs at temperatures $k_\mathrm{B}T \approx 0.8\hbar c/\mathcal{L}$.
The monotonic increase of $\delta_\text{crit}$ with increasing temperature
at large distances is in contrast to the opposite behavior found in the
PFA regime as displayed in Fig.~\ref{fig:pfa_temperature_corrections}.

\subsection{Dipole-plane limit}
Next, we consider the sphere-plane geometry for large distances. The reflection
operator in \eqref{eq:U_PP} thus describes a PEMC plane.
As the translation coefficients \eqref{eq:translation} are diagonal in the
plane-wave basis, the matrix elements are conveniently expressed as
\begin{equation}
\begin{aligned}
U_{P, P'}^{(m)} = \sum_{p, p' = \TE, \TM}&\int \frac{\mathrm{d}\mathbf{k}}{(2\pi)^2}
 e^{-2\kappa\mathcal{L}}
 \langle m, P|\mathbf{k}, p\rangle
\\
&\quad \times
\langle \mathbf{k}, p|\mathcal{R}_{2}|\mathbf{k}, p'\rangle
\langle \mathbf{k}, p'|m, P' \rangle
\end{aligned}
\end{equation}
with $\mathcal{L} = L + R$. 
PEMCs are idealized systems, where the reflection coefficients of a planar surface
neither depend on the frequency nor on the incoming and outgoing wave vector.
In the polarization subspace, the reflection matrix elements 
for a plane characterized by the PEMC parameter $\theta_2$ are thus given by
\cite{Lindell2005, Rode2018}
\begin{equation}
\langle p |\mathcal{R}_2 | p' \rangle = \begin{pmatrix}
\cos(2\theta_2) &  -\sin(2\theta_2)\\
-\sin(2\theta_2) & -\cos(2\theta_2)
\end{pmatrix}\,.
\end{equation}

Together with the basis transformation coefficients mentioned in connection
with \eqref{eq:basis_change}, the Casimir free energy can be calculated analytically
and yields
\begin{equation}
\mathcal{F}_\text{dip-plane} = -\frac{\hbar c}{2 \pi \mathcal{L}}
\left(\frac{R}{\mathcal{L}}\right)^3 \cos(2\delta)
\left[g_{P, P}(\tilde{\tau}) + g_{P, \bar{P}}(\tilde{\tau})\right]\,.
\label{eq:F_dip_plane}
\end{equation}
The functions $g_{P, P}(\tilde{\tau})$ and $g_{P, \bar{P}}(\tilde{\tau})$ account for the
channels conserving and changing polarization during translation,
respectively. They can be found in \cite{IngoldUmrath2015} and are reproduced in
\eqref{eq:dipole_plane} for convenience.

At zero temperature, the functions $g_{P, P}$ and $g_{P,\bar{P}}$ are given by
the numerical factors 15/16 and 3/16, respectively. In the special case $\delta
=0$, our result thus agrees with the one obtained in Ref.~\onlinecite{Emig2008}. In
Fig.~\ref{fig:beyond_pfa}, the dipole-plane result \eqref{eq:F_dip_plane} at
zero temperature is depicted by the dashed lines for various values of
$\delta$. In the high-temperature limit, the polarization-changing contribution
$g_{P, \bar{P}}$ vanishes while $g_{P,P}$ yields $3\tilde{\tau}/8$ which is in agreement
with Ref.~\onlinecite{Canaguier-Durand2010} for $\delta=0$. 

In the large-distance limit, the force changes its sign at the central angle,
\textit{i.e.} $\delta_\text{crit}=\pi/4$, for all temperatures.  The critical
angle only becomes dependent on temperature, if we take terms of the order of
$(R/\mathcal{L})^6$ and higher into account. These terms originate from higher
multipole orders as well as from multiple scatterings between the objects.

\section{High-temperature limit}
\label{sec:high_temp}

In the previous two sections, we have presented analytical results for small
and large distances between the two objects. In order to gain a more complete
understanding of the full range of distances, we consider the high-temperature
limit, where analytical calculations of the Casimir interaction are possible. 
Recent studies \cite{Gelbwaser2022, Ingold2022} showed that already the single-round-trip
contribution provides useful insight into the Casimir interaction over the 
whole distance range between the objects.
In the high-temperature limit \eqref{eq:FT}, the single-round-trip
expression 
\begin{equation}
\mathcal{F}^{(1)}_T = -\frac{k_\mathrm{B}T}{2}\tr\mathcal{M}(0)
\end{equation}
can be calculated analytically. By inserting the reflection matrix elements
\eqref{eq:scat_zerofreq} into \eqref{eq:def_trMr}, we find that the trace can be
expressed in terms of the traces of the round-trip operators for two PEC spheres
($\delta =0$) and the combination of a PEC and a PMC sphere ($\delta= \pi/2$) as
\begin{equation}
\begin{aligned}
\tr\mathcal{M} =  
\cos^2(\delta) \tr\mathcal{M}_\mathrm{PEC-PEC}
 - \sin^2(\delta) \tr\mathcal{M}_\mathrm{PEC-PMC}\,.
\end{aligned}
\label{eq:single_roundtrip}
\end{equation}
The traces for the two limiting cases are given by 
\begin{widetext}
\begin{equation}
\begin{aligned}
\tr\mathcal{M}_\mathrm{PEC-PEC}
 = \frac{R_1 R_2}{\pi^2 \mathcal{L}^2} 
 \int_0^1\mathrm{d}\boldsymbol{t}&
 \int_{-\infty}^\infty \mathrm{d}\mathbf{x} 
 \int_{-\infty}^\infty \mathrm{d}\mathbf{y}
 e^{-(x_1^2 + x_2^2)} e^{-(y_1^2 + y_2^2)} 
\bigg[\left(\cosh(\chi_{12}^{(1)}) -1\right)
 \left(\cosh (\chi_{21}^{(2)}) -1\right) \\
&+
 \left(\cosh(\chi_{12}^{(1)}) - 2t_1\cosh(t_1 \chi_{12}^{(1)})\right)
  \left(\cosh (\chi_{21}^{(2)}) -2t_2 \cosh(t_2 \chi_{21}^{(2)})\right) 
\bigg]
 \end{aligned}
\end{equation}
and 
\begin{equation}
\begin{aligned}
\tr\mathcal{M}_\mathrm{PEC-PMC}
 = \frac{R_1 R_2}{\pi^2 \mathcal{L}^2} 
 \int_0^1\mathrm{d}\boldsymbol{t} &
 \int_{-\infty}^\infty \mathrm{d}\mathbf{x} 
 \int_{-\infty}^\infty \mathrm{d}\mathbf{y}
 e^{-(x_1^2 + x_2^2)} e^{-(y_1^2 + y_2^2)} 
\bigg[\left(\cosh(\chi_{12}^{(1)}) -1\right)
 \left(\cosh (\chi_{21}^{(2)}) -2t_2 \cosh(t_2 \chi_{21}^{(2)})\right)\\ 
& +
 \left(\cosh(\chi_{12}^{(1)}) - 2t_1\cosh(t_1 \chi_{12}^{(1)})\right)
  \left(\cosh (\chi_{21}^{(2)}) -1\right)
\bigg]\,.
 \end{aligned}
\end{equation}
Here, we performed a variable transformation $(x_i, y_i)=\sqrt{k_i \mathcal{L}}(\cos(\varphi_i/2), \sin(\varphi_i/2))$ 
(see Ref.~\onlinecite{Schoger2021} for more details). 
The argument of the hyperbolic cosine reads $\chi_{ij}^{(n)} = 2R_n(x_i x_j + y_i y_j)/\mathcal{L}$. 
The integrals are of Gaussian type and can be calculated by following the approach given in
Ref.~\onlinecite{Schoger2022}, which leads to 
\begin{equation}
\begin{aligned}
\tr\mathcal{M}_\mathrm{PEC-PEC}
= \frac{y}{y^2 - 1}  
+ \frac{1}{z}
+ \frac{z}{6}\log\left(\frac{z^2(y^2-1)}{(yz +1/2)^2}\right)
 - \sum_{\sigma=\pm} \Bigg[\frac{1}{2y + \alpha_\sigma}
 - \frac{1}{6\sqrt{z}}\frac{1}{\alpha_\sigma^{3/2}}
 \log\left(\frac{2y^2 + \alpha_\sigma y -1 + \sqrt{\alpha_\sigma z}}
     {2y^2 + \alpha_\sigma y -1 - \sqrt{\alpha_\sigma z}}\right)\Bigg]
\end{aligned}
\label{eq:single_roundtrip_pec-pec}
\end{equation}
and
\begin{equation}
\begin{aligned}
\tr\mathcal{M}_\mathrm{PEC-PMC}
= \frac{y}{y^2 - 1}  + \frac{z-2y}{2}\log\left(\frac{z^2(y^2-1)}{(yz+1/2)^2}\right) 
- \sum_{\sigma=\pm} \Bigg[\frac{1}{2y + \alpha_\sigma} 
- \frac{1}{2\sqrt{z}} \frac{1}{\alpha_\sigma^{3/2}}
 \log\left(\frac{2y^2 + \alpha_\sigma y -1 + \sqrt{\alpha_\sigma z}}
{2y^2 + \alpha_\sigma y -1 - \sqrt{\alpha_\sigma z}}\right)\Bigg]\,.
\end{aligned}
\label{eq:single_roundtrip_pec-pmc}
\end{equation}
\end{widetext}

In the high-temperature limit, the traces are only functions of the 
geometrical parameters of the system, with
\begin{equation}
y = \frac{\mathcal{L}^2 - R_1^2 - R_2^2}{2R_1R_2} =  1+ x + \frac{u}{2}x^2\,,
\label{eq:def_y}
\end{equation}
where $y$ is a conformally invariant distance scale. The parameters $x$ and
$u$ were introduced in \eqref{eq:def_x} and \eqref{eq:def_u}, respectively. 
Furthermore, we introduced 
\begin{equation}
z = 2y + \alpha_+ + \alpha_-
\end{equation}
with
\begin{equation}
\alpha_\pm = \frac{1-2u \pm \sqrt{1-4u}}{2u}\,.
\label{eq:def_z}
\end{equation}

The result for the sphere-plane geometry follows by taking the limit $u=0$
implying that $\alpha_-$ vanishes while $\alpha_+$ goes to infinity.  The
terms for $\sigma=+$ in \eqref{eq:single_roundtrip_pec-pec} and
\eqref{eq:single_roundtrip_pec-pmc} thus yield zero. By summarizing the
remaining terms one finds that the repulsive magnetoelectric term and the
attractive term become identical in the sphere-plane limit and read
\begin{equation}
  \begin{aligned}
    \tr\mathcal{M}_\mathrm{PEC-PEC}^{u=0} &= \tr\mathcal{M}_\mathrm{PEC-PMC}^{u=0}\\
    &= \frac{y}{y^2-1} - \frac{1}{2y} + \frac{y}{2}\log\left(\frac{y^2-1}{y^2}\right)\,.
  \end{aligned}
\label{eq:single_rountrip_u0}
\end{equation}

\begin{figure}
 \includegraphics[width=0.4\textwidth]{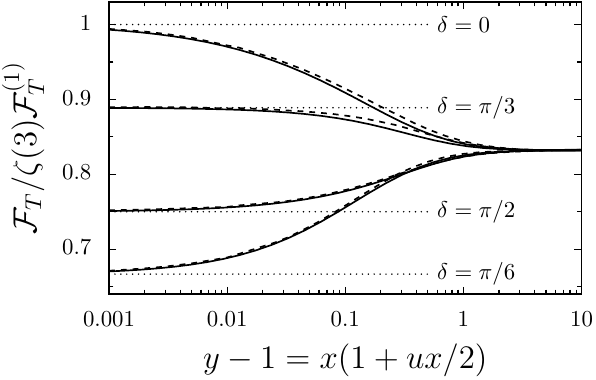}
 \caption{Ratio of the high-temperature free energy $\mathcal{F}_T$ and the
          single-round-trip result $\mathcal{F}_T^{(1)}$ scaled by $\zeta(3)$
          as a function of the conformally invariant distance scale
          $y-1$ for different values of $\delta$ and geometrical
          parameters $u=0$ (solid lines) and $u=1/4$ (dashed lines).
          The dotted lines represent the values $1, 2/3, 8/9$, and
          $3/4$ for the ratio in the PFA limit corresponding to
          $\delta = 0, \pi/6, \pi/3$, and $\pi/2$, respectively.}
 \label{fig:ht_pemc}
\end{figure}

In Fig.~\ref{fig:ht_pemc}, we compare the single-round-trip free energy scaled
by the Apéry constant $\zeta(3)$ with the exact high-temperature result for
$\delta = 0, \pi/6, \pi/3$, and $\pi/2$. At large distances, the ratio
approaches $1/\zeta(3)$ thus confirming that the single-round-trip result becomes
exact. At small distances, it differs from the exact result only by a factor
of order one. In view of \eqref{eq:FT} and \eqref{eq:Fn_PFA}, the ratio at small
distances is given by
\begin{equation}
  \frac{\mathcal{F}_T}{\mathcal{F}_T^{(1)}}
  = \frac{\text{Re}[\text{Li}_3(e^{2i\delta})]}{\cos(2\delta)}
  \ \ \text{for $x\ll 1$.}
\end{equation}
The solid lines for $u=0$ and the dashed lines for $u=1/4$ for the same value
of $\delta$ barely differ, because the data are shown as a function of the
conformally invariant distance scale $y-1$ which itself depends on $u$.

The monotonic behavior of the ratio $\mathcal{F}_T/\mathcal{F}_T^{(1)}$
can be captured by a rational model \cite{Schoger2022b}
\begin{equation}
\Phi_\delta = \prod_{k=1}^n 
\frac{e^{y-1}-1 + \nu_k(\delta)}{e^{y-1}-1 + \mu_k(\delta)}
\label{eq:rational_model}
\end{equation}
with the expansion coefficients obtained from a fit to the numerical results
and given in Table~\ref{tab:expansion_coeffs}. The maximal relative deviation
is of the order of a few per mill as indicated in the last column. The
combination of the rational model with the single-round-trip result thus leads
to a simple approximation of the free energy over the whole distance range.
Note, that for the special case $u=0$ and $\delta=\pi/4$ the single round-trip
does not contribute. Then, the next order in the round-trip expansion
\eqref{eq:def_Fn} needs to be considered which will be discussed in the next
section.

\begin{table}
  \caption{Expansion coefficients for the rational model \eqref{eq:rational_model}
           fitting the ratio $\mathcal{F}_T/\mathcal{F}_T^{(1)}$ for $u=0$
           and $u=1/4$. The last column shows the maximal relative deviation
           $\Delta$ for all values of $u$ and $y$.}
  \label{tab:expansion_coeffs}
  \begin{ruledtabular}
   \begin{tabular}{cccccc}
    $\delta$ & $\nu_1$ & $\nu_2$ & $\mu_1$ & $\mu_2$ & $\Delta\times 10^3$ \\
    \hline
    $0$ & 0.01148 & 0.18511 & 0.01103 & 0.16069 & 4.6 \\
	$\pi/6$ & 0.00020  & 0.07213 & 0.08928 & 0.00021 & 5.1 \\
	$\pi/3$ & 0.17655 & 0.25447 & 0.20505 & 0.20505 & 5.0 \\
	$\pi/2$ & 0.00468 & 0.21056 & 0.23221 & 0.00471 & 1.9 
   \end{tabular}
  \end{ruledtabular}
\end{table}

\section{Sum rule}\label{sec:sum_rule}
Rode et al.\ \cite{Rode2018} observed that the Casimir force
between two parallel PEMC planes at zero temperature
obeys the sum rule
\begin{equation}
\int_0^{\pi/2}d\delta\,F(\delta) = 0
\end{equation}
when integrated over the system parameter $\delta$. As the
distance dependence factors out, the corresponding integral
over the Casimir energy vanishes as well. A similar sum rule
was found for the Casimir energy between a PEMC plate and a
Weyl semimetal \cite{Dudal2023}. 

For two spheres, the sum rule can be expected to still hold
in the PFA regime. At zero temperature, this can indeed be
shown by integrating \eqref{eq:E_PFA} over $\delta$ while at finite
temperatures one can make use of the series expansion of
\eqref{eq:Fn_PFA}. More interesting is the large-distance
limit. Starting from the dipole-dipole result \eqref{eq:F_dip_dip},
one finds
\begin{widetext}
 \begin{equation}
   \int_0^{\pi/2}d\delta F(\delta) = -\frac{\hbar c}{32\mathcal{L}^2}\left(\frac{R_1R_2}{\mathcal{L}^2}\right)^3
   \big[18g(\tilde{\tau})\cosh(\tilde{\tau})+18g(\tilde{\tau})^2
     +14g(\tilde{\tau})^3\cosh(\tilde{\tau})+2g(\tilde{\tau})^4\big(2\cosh^2(\tilde{\tau})+1\big)\big]
 \end{equation}
\end{widetext}
with $g(\tilde{\tau})=\tilde{\tau}/\sinh(\tilde{\tau})$. As a consequence, the sum
rule is violated at zero temperature
\begin{equation}
\int_0^{\pi/2}d\delta F(\delta) = -\frac{7}{4}\hbar c\frac{(R_1R_2)^3}{\mathcal{L}^8}
\end{equation}
as well as in the high-temperature limit
\begin{equation}
\int_0^{\pi/2}d\delta F(\delta) = -\frac{9\pi}{8}k_BT\frac{(R_1R_2)^3}{\mathcal{L}^7}
\label{eq:I_dipdip_ht}
\end{equation}
and for all temperatures in between.

The situation is different when one of the spheres is replaced by a plane. At small
distances, the sum rule is still fulfilled like for the case of two spheres.
However, from the dipole-plane result \eqref{eq:F_dip_plane} one finds that
the sum rule is also satisfied for large distances. Therefore, at large
distances the transition from two spheres to a sphere in front of a plane
is discontinuous.

\begin{figure}
\centering
\includegraphics[scale=0.7]{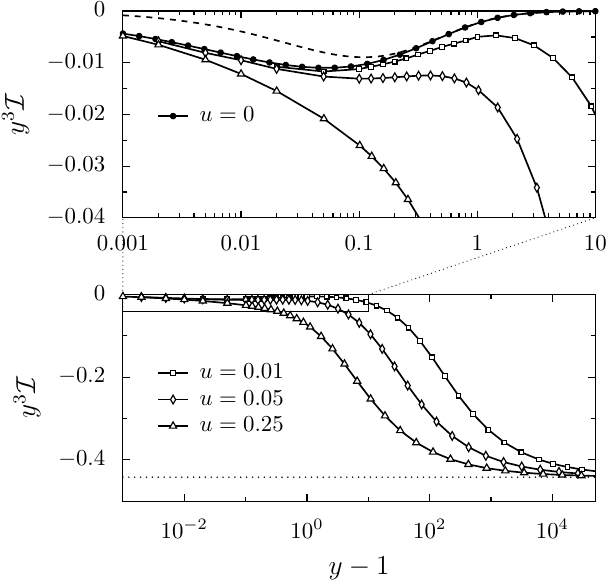}
\caption{Integral of the dimensionless Casimir force \eqref{eq:scaledI_ht}
         in the high-temperature limit scaled by the geometrical factor
         $y^3$. The lower panel depicts the results of the integral for
         the sphere-sphere geometry with $u=0.01$ (open squares), 
         $u=0.05$ (open diamonds) and
         $0.25$ (open triangles). For large separations $y-1\gg 1$,
         the curves converge towards the dipole-dipole result $-9\pi/64$
         as indicated by the dotted line. The upper panel zooms into the
         upper left region of the lower panel and displays, in addition, the
         results for the sphere-plane setup ($u=0$) as filled circles.
         The dashed line corresponds to the values obtained from
         the double-round-trip approximation.}
\label{fig:sumrule}
\end{figure}

In order to obtain a complete picture of the violation of the sum rule in
the high-temperature limit, we present in Fig.~\ref{fig:sumrule} numerical
results as a function of the distance for two equally sized spheres ($u=0.25$,
open triangles), for two spheres of considerably different size ($u=0.05$,
open diamonds and $u=0.01$, open squares), and for a sphere and a plane ($u=0$, filled circles).
We scale the dimensionless integral over the thermal Casimir force
\begin{equation}
 \mathcal{I} = \frac{\mathcal{L}}{k_\text{B}T}\int_0^{\pi/2}\text{d}\delta F_T(\delta)
 \label{eq:scaledI_ht}
\end{equation}
by the geometrical factor $y^3$ which reproduces the large-distance
behavior of the Casimir force for both the sphere-plane and sphere-sphere setup. The value of the integral turns out to always
possess a negative sign, implying that the enclosed area to the left of the
critical angle, for which the force is attractive, is larger than the area to
the right. 

The lower panel of Fig.~\ref{fig:sumrule} shows how the violation of the
sum rule increases from the PFA regime on the left to the dipole-dipole
regime on the right. At large distances, the scaled dimensionless integral
approaches the asymptotic value $-9\pi/64$ as indicated by the dotted line. 
The upper panel represents a zoom into the upper left region of the lower
panel, allowing us to also present numerical data for the plane-sphere setup.
Clearly, in this case the violation of the sum rule is significantly smaller
than for the sphere-sphere setup.

The case of very different sphere radii is particularly interesting. We present
data for $u=0.01$ and $u=0.05$, which according to \eqref{eq:def_u} for such small values
is close to the ratio of the sphere radii. In the small distance regime, the
violation of the sum rule is close to the one obtained for the plane-sphere
setup. For intermediate distances, the violation starts to decrease before
increasing again in order to approach the large-distance result. One thus
observes a non-monotonic behavior of the scaled integral $y^3\mathcal{I}$.

With the integral \eqref{eq:scaledI_ht} for $u=0$ vanishing in the dipole-plane
limit, an analytical description for the plane-sphere setup in the regime
$y\gtrsim 1$ requires to go beyond this limit and even beyond the
single-round-trip approximation \eqref{eq:single_roundtrip}. Accounting 
for two round-trips, the trace over the square of the round-trip matrix
yields
\begin{equation}
\begin{aligned}
\tr\mathcal{M}^2_{u=0} =& \cos^2(2\delta) \tr\mathcal{M}^2_\mathrm{PEC-PEC} \\
& - \sin^2(2\delta) \tr\mathcal{M}^2_\mathrm{PEC-PMC}\,.
\end{aligned}
\label{eq:trM2}
\end{equation}
The double-round-trip expression for the respective limiting cases of $\delta =
0$ and $\pi/2$ can be obtained from the single-round-trip results
\eqref{eq:single_roundtrip_pec-pec} and \eqref{eq:single_roundtrip_pec-pmc} for
two spheres with the same radii \cite{Bulgac2006} by replacing $y$ by $2y^2-1$.
The explicit expressions are given in Appendix~\ref{sec:double_roundtrip}.
As the two traces in \eqref{eq:trM2} differ, the double-round-trip
expression depicted by the dashed curve in the upper panel of 
Fig.~\ref{fig:sumrule} describes the leading violation of the sum rule 
at intermediate and large distances between sphere and plane.

\section{Discussion} \label{sec:discussion}
In the previous sections, we have examined the Casimir interaction for small
and large distances as well as for low and high temperatures. The Casimir force
was found to vanish for a critical material parameter $\delta_\text{crit}$
which depends on the geometry and the temperature. In the following, we will
discuss these results by focussing on the existence of an equilibrium position
and its dependence on the temperature.  While stable equilibrium
positions were ruled out for reciprocal objects in vacuum \cite{Rahi2010}, they
are possible for non-reciprocal materials \cite{Gelbwaser2022} as we shall see. 

\begin{figure}
\includegraphics[width=0.3\textwidth]{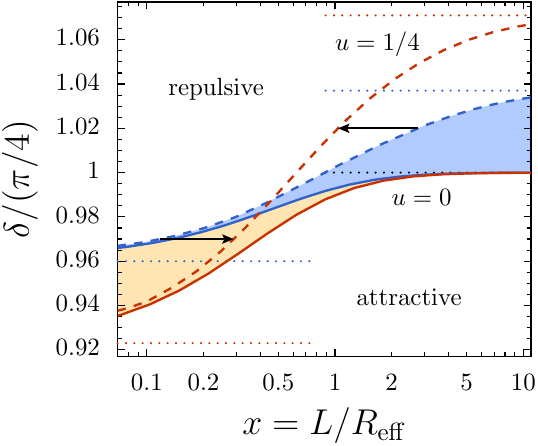}
\caption{Curves of vanishing Casimir force separating regions of attractive and
         repulsive force are shown as a function of the aspect ratio $x$ and the 
         material parameter $\delta$. Solid and dashed lines correspond to the sphere-plane geometry
         ($u=0$) and two equally sized spheres ($u=1/4$), respectively, while blue
         and red curves indicate the zero-temperature case and the high-temperature
         limit, respectively. The region covering all temperatures for $u=0$ is shaded
         in yellow while the region covering all aspect ratios in the zero-temperature
         limit are shaded in blue. The dotted lines indicate the asymptotic values for small
         and large distances. The arrows mark the opposite change of the equilibrium position
         with increasing temperature for $u=1/4$ for different values of $\delta$.}
\label{fig:xvsdelta}
\end{figure}

In Fig.~\ref{fig:xvsdelta}, we show the curves of vanishing Casimir force as a
function of the aspect ratio $x$ and the material parameter $\delta$ for the
sphere-plane geometry ($u=0$, solid lines) and two equally sized spheres
($u=1/4$, dashed lines).  The depicted curves delimit a region where curves for
intermediate values of $u$ can be found as is exemplified by the blue shaded area
related to the case of zero temperature. The curves in Fig.~\ref{fig:xvsdelta}
separate regions of repulsive Casimir force at smaller distances and attractive
Casimir force at larger distances. They are therefore associated with stable
equilibrium positions. For both geometries, blue and red curves depict the
zero- and high-temperature limit, respectively. Curves for finite temperatures
lie in between as indicated by the yellow shaded area for $u=0$. All curves
increase monotonically in $\delta$ with increasing distance and converge
towards the critical angles computed earlier and depicted by the dotted lines
for small and large distances.

In our discussion of the PFA corrections in Sec.~\ref{sec:pfa_corrections}, we
observed that curvature effects become more important as the distance between
the objects increases. The results presented in Fig.~\ref{fig:xvsdelta} confirm
that the variation in the critical angle is smaller for a sphere in front of a
plane as compared to two equally sized spheres. From a practical point of view,
however, we are not so much interested in the effect of a variation of sphere
radii which is difficult to realize, but rather in the effect of the more
easily controllable temperature on the existence and variation of the
equilibrium positions. The geometrical parameter $u$ and the material parameter
$\delta$, are thus fixed in the following.

\begin{figure}
\includegraphics[width=0.4\textwidth]{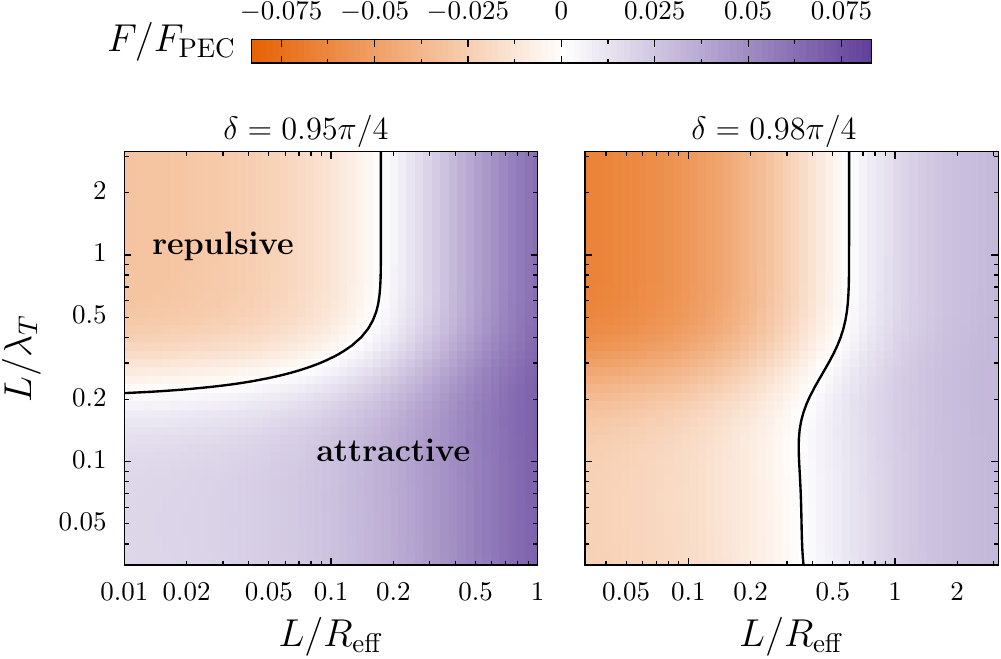}
\caption{Casimir force relative to the Casimir force for two perfect reflectors
         as a function of the effective distance and temperature for the
         sphere-plane geometry ($u=0$) with $\delta=0.95\pi/4$ (left) 
         and $0.98\pi/4$ (right). Negative (positive) values correspond to
         repulsion (attraction). The Casimir force vanishes along the solid line.}
\label{fig:fu0delta}
\end{figure}

We start with the sphere-plane geometry. According to the blue and red solid
lines depicted in Fig.~\ref{fig:xvsdelta}, the equilibrium distance increases
with increasing temperature if $0.96\pi/4\lesssim\delta < \pi/4$. Equilibrium
positions do not exist for any temperature if the material parameter $\delta$
exceeds $\pi/4$. In the regime $0.92\pi/4\lesssim\delta \lesssim 0.96\pi/4$,
the Casimir force can only vanish for not too low temperatures. A clearer
picture of the temperature dependence can be obtained from
Fig.~\ref{fig:fu0delta}, where the Casimir force relative to the Casimir force
for two perfect reflectors ($\delta=0$) is shown as a function of the aspect
ratio and the temperature. The chosen material parameters $\delta = 0.95\pi/4$
and $0.98\pi/4$ lie in the second and first range of material parameters,
respectively, and correspond to two distinct scenarios. For the first value, an
equilibrium position only exists for temperatures above $L/\lambda_T \approx
0.2$. Below this threshold, the force is always attractive. For
$\delta=0.98\pi/4$, on the other hand, there exists an equilibrium distance for
all temperatures, which increases until the high-temperature distance
depicted by the solid red line in Fig.~\ref{fig:xvsdelta} is reached.

\begin{figure}
\includegraphics[width=0.4\textwidth]{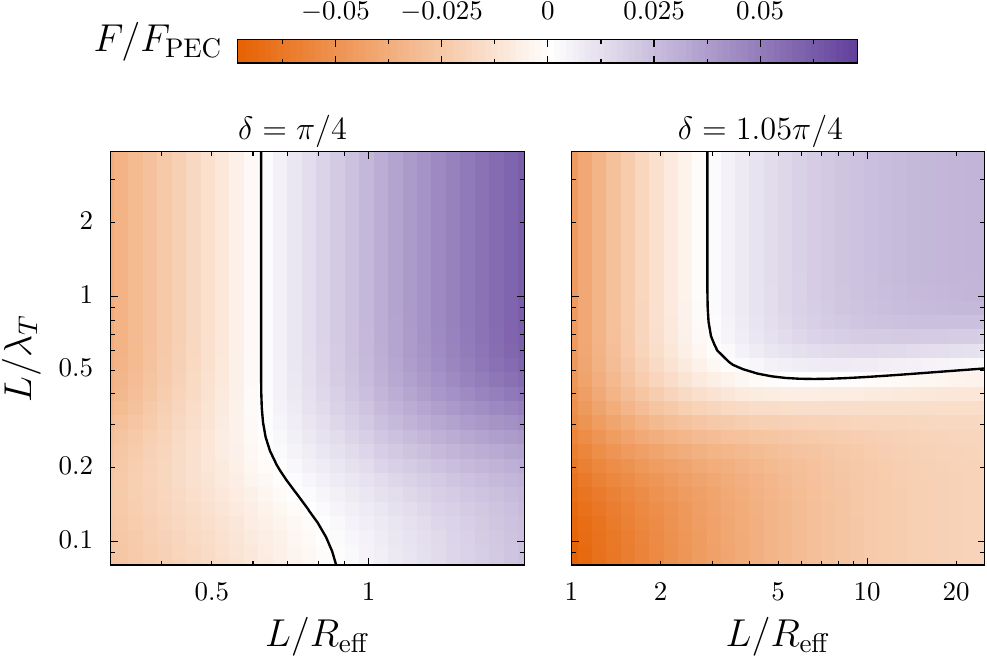}
\caption{Casimir force relative to the Casimir force for two perfect reflectors
         as a function of the effective distance and temperature for two
         equally-sized spheres ($u=1/4$) with $\delta=\pi/4$ (left) 
         and $1.05\pi/4$ (right). Negative (positive) values correspond to
         repulsion (attraction). The Casimir force vanishes along the solid line.}
\label{fig:fu025delta}
\end{figure}

Turning to two equally-sized spheres ($u=1/4$), we first note that according to
Fig.~\ref{fig:xvsdelta} there exists a critical material parameter of
approximately $0.99\pi/4$ where the equilibrium distance is independent of
temperature. Below this value, the sphere-sphere geometry behaves qualitatively
like the sphere-plane geometry discussed before. As the lower black arrow
illustrates, the equilibrium distance is pushed to larger distances as
temperature increases. The behavior changes for material parameters in the
range $0.99\pi/4 \lesssim\delta\lesssim 1.04 \pi/4$. There, the equilibrium
distance decreases as the temperature increases, as indicated by the upper
arrow and shown in Fig.~\ref{fig:fu025delta} on the left for the specific material
parameter $\pi/4$. For even larger values of $\delta$ like for the case 
$\delta=1.05\pi/4$ shown in the right panel of Fig.~\ref{fig:fu025delta},
the Casimir force vanishes only for sufficiently large temperatures.

\section{Conclusions}\label{sec:conclusions}

In this paper, we have extended the study of the Casimir interaction between
perfect electromagnetic conductors to two spherical objects at finite
temperatures. The Casimir interaction depends on a material parameter $\delta$
which tunes the magneto-electric response and thus allows for a transition
between an attractive and a repulsive Casimir force. 

We found that the transition point $\delta_\mathrm{crit}$ depends on the
geometric parameters as well as on temperature. For values of $\delta$ around
$\pi/4$, there exist sphere-sphere and sphere-plane setups where an
equilibrium configuration is possible. The existence of equilibria requires the
use of non-reciprocal materials to which PEMCs belong. For practical purposes,
it is of particular interest that the equilibrium distance depends on the
temperature, thus offering a scenario where the sign of the Casimir force can
be controlled. Our theoretical study of the idealized system of perfect
electromagnetic conductors may serve as a guide to explore more realistic
materials for a geometry commonly employed in Casimir experiments.

\begin{acknowledgments}
  The authors are grateful to Paulo A. Maia Neto, Serge Reynaud and Benjamin Spreng for
  insightful discussions.
\end{acknowledgments}

\appendix

\section{Scattering coefficients for a biisotropic sphere} \label{sec:mie_coef}
We consider the scattering at a general biisotropic sphere of radii $R$ in vacuum 
described by the matrix element
\begin{equation}
\langle \ell, m , P, \mathrm{out}| \mathcal{R} | \ell', m', P', \mathrm{reg}\rangle
= - i^{P'-P} R_{P, P'} \delta_{\ell, \ell'} \delta_{m, m'}\,.
\end{equation}
While the angular momentum variables $\ell, \ell', m,$ and $m'$ are conserved during
the scattering process, this is not the case for the polarization.
The coefficients $R_{P, P'}$ account for the Mie coefficients as introduced in 
\eqref{eq:R_multipole}. The polarization-conserving scattering coefficients 
at imaginary frequencies 
$\xi = -i\omega$ are given by
\begin{align}
R_{\E\E}(i\tilde{\xi}) &= C_\ell(\tilde{\xi})
\frac{W^R_\ell(\tilde{\xi}) A^L_\ell(\tilde{\xi}) + W^L_\ell(\tilde{\xi}) A^R_\ell(\tilde{\xi})}
{V^R_\ell(\tilde{\xi}) W^L_\ell(\tilde{\xi}) + V^L_\ell(\tilde{\xi}) W^R_\ell(\tilde{\xi})}
\end{align}
\begin{align}
R_{\M\M}(i\tilde{\xi}) &= C_\ell(\tilde{\xi})
\frac{V^R_\ell(\tilde{\xi}) B^L_\ell(\tilde{\xi}) + V^L_\ell(\tilde{\xi}) B^R_\ell(\tilde{\xi})}
{V^R_\ell(\tilde{\xi}) W^L_\ell(\tilde{\xi}) + V^L_\ell(\tilde{\xi}) W^R_\ell(\tilde{\xi})}
\end{align}
where we introduced for convenience 
\begin{equation}
C_\ell(\tilde{\xi})  = (-1)^\ell \frac{\pi}{2} \frac{I_{\ell+1/2} (\tilde{\xi})}{K_{\ell+1/2}(\tilde{\xi})} \,.
\end{equation}
Here , $I_{\ell+1/2}(\tilde{\xi})$ and $K_{\ell+1/2}(\tilde{\xi})$ are modified
Bessel functions of first and second kind, respectively, of fractional
order. The dimensionless frequency is given by
\begin{equation}
\tilde{\xi} = \frac{\xi R}{c}\,.
\end{equation}
Moreover, we adopted the notation from Ref.~\onlinecite{BohrenHuffman2004} and defined 
the following auxiliary variables
\begin{equation}
\begin{aligned}
A^{L, R}_\ell (\tilde{\xi}) &=  \{I,\tilde{\xi}\} - m_{\mp}\{I, \tilde{\xi} m_{L, R}\} \\
B^{L, R}_\ell (\tilde{\xi}) &=  m_{\mp}\{I,\tilde{\xi}\} - \{I, \tilde{\xi}  m_{L, R}\} \\
V^{L, R}_\ell (\tilde{\xi}) &=  m_\mp \{I, \tilde{\xi} m_{L, R}\} - \{K,\tilde{\xi}\} \\
W^{L, R}_\ell (\tilde{\xi}) &= \{I,\tilde{\xi} m_{L, R}\} - m_{\mp}\{K,\tilde{\xi}\}
\end{aligned}
\end{equation}
where following Ref.~\onlinecite{Nussenzveig1992} we introduced the notation
\begin{equation}
\{\mathcal{I}, z\} = \frac{\mathcal{I}'_{\ell +1/2}(z)}{\mathcal{I}_{\ell +1/2}(z)} 
	+ \frac{1}{2z}
\end{equation}
with $\mathcal{I} = I, K$. 
The relative refractive indices $m_{L, R}$ for left- and right-polarized light 
are given by
\begin{equation}
m_{L, R} = 
\sqrt{\epsilon\mu - \left((\beta+\alpha)/2\right)^2} \pm i (\beta-\alpha)/2\,.
\end{equation}
The indices $m_\pm$ account for the relative impedances
\begin{equation}
m_\pm 
= \frac{1}{\epsilon}
\left[\sqrt{\epsilon\mu - \left((\beta +\alpha)/2\right)^2} 
	\mp i (\beta+\alpha)/2\right]\,.
\end{equation}
It can easily be verified that the Mie coefficients for the electric and magnetic modes 
yield the known results in the isotropic limit, where $\alpha = \beta = 0$. 
The polarization mixing coefficients are defined as 
\begin{equation}
\begin{aligned}
R_{\M\E}(i\tilde{\xi}) =& iC_\ell(\tilde{\xi})\left[\{I,\tilde{\xi}\} - \{K,\tilde{\xi}\}\right] \\
	&\times\frac{m_-\{I,\tilde{\xi} m_R\} - m_+\{I,\tilde{\xi} m_L\}}
		{V^R_\ell(\tilde{\xi}) W^L_\ell(\tilde{\xi}) + V^L_\ell(\tilde{\xi}) W^R_\ell(\tilde{\xi})}
\end{aligned}
\end{equation}
\begin{equation}
\begin{aligned}
R_{\E\M}(i\tilde{\xi}) =& iC_\ell\left[\{I,\tilde{\xi}\} - \{K,\tilde{\xi}\}\right] \\
&\times\frac{m_-\{I,\tilde{\xi} m_L\} - m_+\{I,\tilde{\xi} m_R\}  
		   }{V^R_\ell(\tilde{\xi}) W^L_\ell(\tilde{\xi}) + V^L_\ell(\tilde{\xi}) W^R_\ell(\tilde{\xi})}\,.
\end{aligned}
\end{equation}

In the PEMC limit, we find according to \eqref{eq:def_param_pemc} that $m_{L, R}$ goes to infinity 
while $m_\pm = \mp i \tan(\theta)$ with $\theta$ taking values
between 0 and $\pi/2$ accounting for a perfect electric 
and perfect magnetic conductor, respectively. 
The polarization conserving Mie coefficients thus yield
\begin{equation}
R_{\E\E} = - C_\ell 
\left[\cos^2(\theta) \frac{\{I, \tilde{\xi}\}}{\{K, \tilde{\xi}\}} + \sin^2(\theta)\right]
\label{eq:al_pemc}
\end{equation}
\begin{equation}
R_{\M\M} = -C_\ell 
\left[\cos^2(\theta)  + \sin^2(\theta) \frac{\{I, \tilde{\xi}\}}{\{K, \tilde{\xi}\}}\right]\,.
\end{equation}
The polarization-mixing coefficients are the same in the PEMC limit and they are given by
\begin{align}
R_{\E\M} = R_{\M\E} =  -C_\ell 
\left[\frac{\{I, \tilde{\xi}\}}{\{K, \tilde{\xi}\}} -1 \right]\frac{\sin(2\theta)}{2}\,.
\label{eq:dl_pemc}
\end{align}
It can now easily be shown that the Mie coefficients for a PEMC sphere can 
be obtained  from the coefficients for a PEC by performing the transformation 
given in Eq.~\eqref{eq:R_PEMC}. 

The reflection matrix elements in the plane-wave basis are expressed in terms
of the amplitude scattering matrix \eqref{eq:refl_coef_planewave}. The
scattering matrix connects an incident plane wave with a plane wave
in the far-field of the object and is given by
\begin{equation}
S_{p,p'} = \sum_{\ell=1}^\infty \frac{2\ell+1}{\ell(\ell+1)} 
(R_{P,P'}\tau_\ell(z) + (-1)^{p-p'} R_{\bar{P}, \bar{P'}}\pi_\ell(z)) 
\label{eq:scattering_ampl}
\end{equation}
where we identify $p=1 (2)$ with $\TM (\TE)$ and $P=1 (2)$ with $\E (\M)$. The angular functions
$\tau_\ell(z)$ and $\pi_\ell(z)$ defined in Ref.~\onlinecite{BohrenHuffman2004}
depend on the in- and outgoing wave vectors
\begin{equation}
z = - \frac{c^2}{\xi^2}\left(k k' \cos(\varphi - \varphi')  +
\kappa\kappa'
\right)\,.
\end{equation}

The term proportional to $\pi_\ell$ can be neglected in the zero-frequency limit. 
The amplitude scattering matrix for a PEMC sphere can thus be expressed
in terms of the scattering matrix for a PEC sphere
\begin{equation}
\mathbf{S}_\mathrm{PEMC} = \mathbf{D}\mathbf{S}_\mathrm{PEC}\mathbf{D}^{-1}
\end{equation}
with the transformation matrix defined in \eqref{eq:def_D}.
The matrix elements of the PEC scattering matrix are 
\begin{equation}
\begin{aligned}
(\mathbf{S}_\text{PEC})_{\TM, \TM} 
&= \frac{\xi R}{c}\left[\cosh(\chi) - 1\right]
\\
(\mathbf{S}_\text{PEC})_{\TE, \TE}
&=  -\frac{\xi R}{c}
 \left[\cosh(\chi) - 2\int_0^1 t \cosh(t\chi)\mathrm{d} t \right]
\end{aligned}
\label{eq:scat_ampl_zerofreq}
\end{equation}
where the argument of the hyperbolic cosine is defined as
\begin{equation}
\chi =  2R \sqrt{k k'} \cos\left(\frac{\varphi - \varphi'}{2}\right)\,.
\end{equation}

\section{Dipole limit}

The Casimir free energy for large distances can be expressed in terms
of dimensionless functions.
For the sphere-sphere geometry the functions introduced in \eqref{eq:F_dip_dip}
are given by 
\begin{equation}
\begin{aligned}
f_{P, P}(\tilde{\tau}) 
 &= \frac{5}{8}\Big[
	6g(\tilde{\tau}) \cosh(\tilde{\tau}) + 6g^2(\tilde{\tau}) + 5g^3(\tilde{\tau})\cosh(\tilde{\tau}) 
	\\  
	&\hspace{2em}
	+ g^4(\tilde{\tau}) (1+ 2\cosh^2(\tilde{\tau}))  \\
	& \hspace{2em}
	+ g^5(\tilde{\tau})\cosh(\tilde{\tau})(2+\cosh^2(\tilde{\tau}))
\Big]
\\
f_{P, \bar{P}}(\tilde{\tau}) 
&= \frac{1}{2}\Big[
	g^3(\tilde{\tau})\cosh(\tilde{\tau}) + g^4(\tilde{\tau}) (1+ 2\cosh^2(\tilde{\tau})) 
	\\ 
	&\hspace{2em} + g^5(\tilde{\tau})\cosh(\tilde{\tau})(2+\cosh^2(\tilde{\tau}))
\Big]
\end{aligned}
\label{eq:dipole_dipole}
\end{equation}
where $g(\tilde{\tau}) = \tilde{\tau}/\sinh(\tilde{\tau})$ with $\tilde{\tau} = 2\pi \mathcal{L}/\lambda_T$. 
The functions can also be obtained by combining Eqs.~(42)--(46) in
Ref.~\onlinecite{IngoldUmrath2015}.

The large-distance behavior for the sphere-plane geometry \eqref{eq:F_dip_plane}
is characterized by the following functions
\begin{equation}
\begin{aligned}
g_{P, P}(\tilde{\tau}) & = \frac{3}{16}\left[2g(\tilde{\tau})\cosh(\tilde{\tau}) + 2g^2(\tilde{\tau})
+ g^3(\tilde{\tau})\cosh(\tilde{\tau})
\right] 
\\
g_{P, \bar{P}}(\tilde{\tau}) &= \frac{3}{16} g^3(\tilde{\tau})\cosh(\tilde{\tau})
\end{aligned}
\label{eq:dipole_plane}
\end{equation}
which can be found in Eqs.~(20)--(22) of Ref.~\onlinecite{IngoldUmrath2015}.

\section{Double round-trip in the high-temperature limit}
\label{sec:double_roundtrip}

In this appendix, we give the double-round-trip expression for the sphere-plane geometry in the high-temperature limit.
The expression can be obtained from the single-round-trip result for two equally-sized spheres, due to the reflection
symmetry of the two-sphere setup with respect to a plane perpendicular to the $z$-axis.
The surface-to-surface distance between the spheres can as a result be seen as twice the distance for the sphere-plane geometry.
Hence, after setting $u=1/4$ in \eqref{eq:single_roundtrip_pec-pec} and \eqref{eq:single_roundtrip_pec-pmc} and replacing
$y$ with $2y^2-1$, we obtain the following double-round-trip expressions
\begin{widetext}
\begin{equation}
\begin{aligned}
\tr\mathcal{M}^2_\mathrm{PEC-PEC} =  
\frac{2y^2-1}{4y^2(y^2-1)}  + \frac{1}{4y^2}
+ \frac{2y^2}{3} \log\left(\frac{ y^6 (y^2-1)}{(y^2-1/4)^4}\right) 
 - \frac{2}{4y^2 -1}
+ \frac{1}{6y}\log\left(\frac{4y^3-3y +1}{4y^3 -3y -1}\right)
\end{aligned}
\end{equation}
\begin{equation}
\begin{aligned}
\tr\mathcal{M}^2_\mathrm{PEC-PMC} = & 
\frac{2y^2-1}{4y^2(y^2-1)} 
+ \log\left(\frac{ y^6 (y^2-1)}{(y^2-1/4)^4}\right) 
 - \frac{2}{4y^2 -1}
+ \frac{1}{2y}\log\left(\frac{4y^3-3y +1}{4y^3 -3y -1}\right) 
\end{aligned}
\end{equation}
where $y= 1+ L/R$ with the surface-to-surface distance $L$ between a plane and a sphere with radius $R$.
\end{widetext}

\bibliographystyle{apsrev}
\bibliography{pemc}

\end{document}